\documentclass[prd,aps,preprint,tightenlines,nofootinbib,superscriptaddress]{revtex4}
\usepackage{epsfig}
\usepackage{amsmath}
\input{epsf}
\usepackage{psfrag}
\usepackage{xcolor}
\usepackage{slashed}
\usepackage{hyperref}
\usepackage{ulem}
\usepackage{CJKutf8}

\begin{document}


\vspace*{2cm}
\title{Accessing the gluon GTMD $\boldsymbol{F_{1,4}}$ in exclusive $\boldsymbol{\pi^0}$ production \\ in $\boldsymbol{ep}$ collisions }

\author{Shohini Bhattacharya}
\affiliation{RIKEN BNL Research Center, Brookhaven National Laboratory, Upton, NY 11973, USA}

\author{Duxin Zheng}
\affiliation{Shandong Institute of Advanced Technology, Jinan, Shandong, 250100, China}

\author{Jian Zhou}
\affiliation{School of Physics and Key Laboratory of Particle Physics and Particle Irradiation (MOE), Shandong University, QingDao, Shandong, 266237, China}

\begin{abstract}
We demonstrate  that the longitudinal single target-spin asymmetry in exclusive $\pi^0$ production in $ep$ collisions can give access to the imaginary part of the gluon generalized transverse momentum distribution (GTMD) $F_{1,4}$. Such a longitudinal spin asymmetry that results from the Coulomb-nuclear interference effect, leads to a characteristic azimuthal angular correlation of $\sin 2\phi$, where $\phi$ is the azimuthal angle between the scattered lepton transverse momentum and the recoil proton's transverse momentum. We also present a numerical estimate of the asymmetry for  the kinematics accessible at EIC and EicC.

\end{abstract}

\maketitle

\date{\today}

\section{Introduction} \label{s:intro}
The study of nucleon spin structure triggered by the discovery of ``spin crisis"~\cite{EuropeanMuon:1987isl}  has grown into a fascinating research area during the past three decades.  One of the main focus of the field is to address the spin decomposition of nucleon. According to the Jaffe-Manohar decomposition~\cite{Jaffe:1989jz},  the proton spin receives contributions from four different sources,
 \begin{eqnarray}
\frac{1}{2}=\frac{1}{2}\Delta \Sigma+\Delta G+L_q+L_g
\end{eqnarray}
where the quark spin contribution $\Delta \Sigma \sim 0.3$ obtained from the integral of the quark helicity distribution is relatively well constrained. The gluon spin contribution is likely to be sizable as well~\cite{STAR:2014wox,deFlorian:2014yva,Nocera:2014gqa,STAR:2021mqa}, though  the  uncertainties  about the gluon  helicity distribution at small $x$ are huge. The gluon spin contribution is expected to be precisely determined with the experimental data from the future Electron-Ion Collider (EIC)~\cite{Boer:2011fh,AbdulKhalek:2021gbh}.

Apart from the gluon helicity distribution at small $x$, the current major focus is the orbital angular momentum (OAM) from quarks and gluons, which could help  deepen our understanding of the partonic structure of nucleon and the associated QCD dynamics ~\cite{Boussarie:2019icw,Guo:2021aik,Kovchegov:2019rrz,Engelhardt:2020qtg}. However, so far very little is known about the  orbital angular momentum of partons.  From both theoretical and experimental point of view, it remains very challenging to extract Jaffe-Manohar type (or canonical) parton OAM  from  high energy scattering experiment. 
All the previous proposals for probing  gluon OAM rely on its connection with the Wigner distribution~\cite{Belitsky:2003nz}, or equivalently, the generalized transverse momentum distribution function (GTMD)~\cite{Lorce:2011kd,Hatta:2011ku,Lorce:2011ni}, 
\begin{eqnarray} 
xL_{g}(x,\xi)=-\int d^2 k_\perp \frac{ 
 k_\perp^2}{M^2} F_{1,4}^{g}(x, k_\perp,\xi, \Delta_\perp=0)
\end{eqnarray}
where the relevant kinematic variables will be specified below. The parton OAM can be reconstructed with the integral of the $x$ dependent  OAM  distribution: $L_{g}=\int_0^1 dx L_{g}(x,\xi=0)$. A considerable amount of  theoretical efforts have been made to explore the experimental signals of the GTMD $F_{1,4}$ in the recent years. The single and the double longitudinal spin asymmetry in exclusive di-jet production in $ep$ collisions was shown to give access to the ``Compton form factor" that involves $F_{1,4}$~\cite{Ji:2016jgn,Hatta:2016aoc,Bhattacharya:2022vvo}. It is also feasible  to directly measure $F_{1,4}$ among many other  GTMDs in the exclusive double Drell-Yan or quarkonium production in hadron collisions~\cite{Bhattacharya:2017bvs,Bhattacharya:2018lgm,Boussarie:2018zwg}.  Note that the kinetic OAM by Ji~\cite{Ji:1996ek} can not be reconstructed from $F_{1,4}$ measured in high energy scatterings as it always involves a light cone gauge link. Another attempt to find an observable for probing OAM can be found in Ref.~\cite{Courtoy:2013oaa}.

In this work we propose a new  observable to extract gluon GTMD $F_{1,4}$. We consider single spin asymmetry in the  exclusive $\pi^0$ production in proton-electron collisions $ep \rightarrow  e' p' \pi^0$ where the proton is longitudinally polarized and the electron is unpolarized.  $\pi^0$ can be exclusively produced off a longitudinally  polarized proton either through two gluon exchange described by $F_{1,4}$ or though a quasi-real photon exchange. The latter is commonly referred to as the Primakoff process.  We show that  an azimuthal angular correlation: $\lambda (\Delta_\perp \cdot l_\perp')(\Delta_\perp \times l_\perp')$  arises from the interference amplitudes of the mentioned two processes, where $\Delta_\perp$, $l_\perp'$ are the nucleon recoiled transverse momentum and the outgoing lepton transverse momentum respectively, and $\lambda$ denotes proton helicity. $F_{1,4}$ thus can be extracted from this characteristic azimuthal angular correlation of $\sin 2\phi $.

The paper is organized as follows. In Sec.II, we present a detailed analytic derivation of the spin dependent cross section following a brief review of the unpolarized cross section calculation. It actually came as a big surprise that, apart from the Primakoff process, the dominant contribution to the unpolarized exclusive $\pi^0$ production in the forward limit was found to be the spin dependent odderon (or the gluon Sivers function)~\cite{Boussarie:2019vmk}. We perform the numerical estimations of the spin asymmetry with some toy model inputs.  Finally, the paper is summarized in Sec.III.

\section{Probing the gluon GTMD $\boldsymbol{F_{1,4}}$ in exclusive $\boldsymbol{\pi^0}$ production}
\label{s:main}
In this section, we compute the gluon GTMD $F_{1,4}$ contribution to the longitudinal target-spin asymmetry in exclusive $\pi^0$ production in $ep$ collisions. The kinematics of the process under consideration is specified as follows,
 \begin{eqnarray}
 e(l)+p(p,\lambda) \longrightarrow \pi^0(l_\pi)+e(l')+p(p',\lambda') \, .
\end{eqnarray}
 It is convenient to define kinemaitc variables $Q=-q^2=-(l-l')^2$, $t=(p'-p)^2$, $y=p \cdot q/p \cdot l$, and the squared photon-nucleon center of mass energy $W^2=(p+q)^2$. We neglect pion mass in our calculation $l_\pi^2=0$.  The skewness variable is given by,
$\xi = (p^+ - p'^+) / (p^+ + p'^+) = - \Delta^+ / (2 P^+)=\frac{x_B}{2-x_B}$
with $x_B=Q^2/2p\cdot q$. Here `$+$' indicates the light-cone plus component which is commonly defined. The momentum transfer squared can be further approximated as $t=-\Delta_\perp^2$ where $\Delta_\perp$ is the nucleon recoiled transverse momentum. 

The exclusive $\pi^0$ production in $ep$  collisions is conventionally used to get a handle on the leading twist helicity flip quark  GPDs when the proton target is transversely polarized~\cite{Frankfurt:1999fp}. If proton target is unpolarized, the cross section of the exclusive $\pi^0$ production in the forward limit has long been thought to vanish. However, it has been found in Ref.~\cite{Boussarie:2019vmk} that the gluon Sivers function plays a surprising role in unpolarized processes, and was shown to  give the dominant contribution to the unpolarized exclusive pion production in the forward limit. This might be one of the most promising channels to search for the elusive odderon as the gluon Sivers function is related to the $k_\perp$ moment of the spin dependent odderon~\cite{Boer:2015pni,Zhou:2013gsa}.  In this work, we consider the longitudinally polarized target case which allows us to probe the $k_\perp$ moment of the GTMD $F_{1,4}$.

 Before presenting the calculation details, let us first recall the operator definition of the GTMDs. We start with introducing the matrix element definition of the GTMD $F_{1,4}$ spin-$\frac{1}{2}$ target~\cite{Meissner:2009ww,Lorce:2013pza},
\begin{eqnarray} 
&&\!\!\!\!\!\!\!\!\!\! \frac{1}{P^+} \int \frac{dz^-  d^2\vec{z}_\perp}{(2\pi)^3}  e^{i k \cdot z} 
\langle p', \lambda' |  {\rm Tr} \! \left [ F_a^{+ \mu}(- \tfrac{z}{2}) \, U^\dag (- \tfrac{z}{2}, \tfrac{z}{2}) \,  F_b^{+ \mu}(\tfrac{z}{2})U (- \tfrac{z}{2}, \tfrac{z}{2}) \right ]  | p, \lambda \rangle \Big|_{z^+ = 0} 
\nonumber \\
& = & \frac{1}{2M} \, \bar{u}(p',\lambda') \bigg[ 
F_{1,1}^g + \frac{i  \sigma^{i+}  k_\perp^i}{P^+} \, F_{1,2}^g + \frac{i \, \sigma^{i+} \Delta_\perp^i}{P^+} \, F_{1,3}^g 
+ \frac{i \sigma^{ij} k_{\perp}^i \Delta_{\perp}^j}{M^2} \, F_{1,4}^g  \bigg] u(p,\lambda)
\nonumber \\
& \approx & \frac{1}{\sqrt{1 - \xi^2}} \bigg\{ 
  \delta_{\lambda,\lambda'}   F_{1,1}^g
+ \frac{i \varepsilon_\perp^{ij} k_{\perp}^i \Delta_{\perp}^j}{M^2}  \lambda  \delta_{\lambda,\lambda'}    F_{1,4}^g +...\bigg\} \,,
\label{e:GTMD_unpol}
\end{eqnarray}
where the gluons are represented by components of the field strength tensor $F_a^{\mu \nu}$, where `$a$' is a color index.
Note that we have limited ourselves to the case of leading twist --- $i, j$ are transverse indices. 
The Wilson line $U$ and $U^\dag$ in  the fundamental representation form a closed loop gauge link.
The average longitudinal and transverse gluon momenta are denoted by $x$ and $k_\perp$, respectively.
In the third line we have explicitly worked out spinor products and only kept the leading power terms in the near forward region $\Delta_\perp \rightarrow 0$. The contributions associated with GTMDs $F_{1,2}$ and $F_{1,3}$ are suppressed as well. 
A number of model calculations of GTMDs is available by now~\cite{Meissner:2008ay, Meissner:2009ww,Lorce:2011kd,Kanazawa:2014nha,Mukherjee:2014nya,Hagiwara:2016kam,Zhou:2016rnt,Courtoy:2016des,Boer:2023mip,Boer:2021upt,Tan:2023kbl,Xu:2022abw,Ojha:2022fls}.

\begin{figure}
    \centering    
    \includegraphics[scale=0.6]{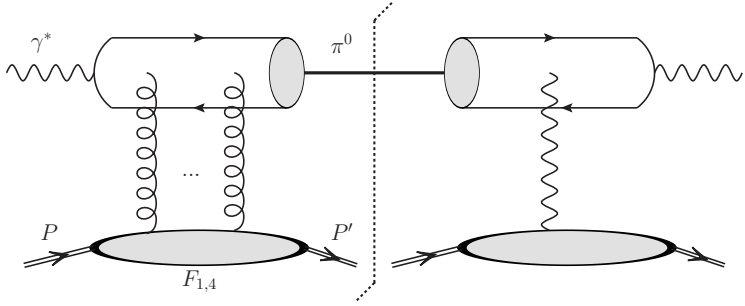}
    \caption{Coulomb-nuclear interference contribution to the exclusive $\pi^0$ production in $ep$ collisions. The imaginary phase of $F_{1,4}$ could be generated through the multiple gluon exchange encoded in the gauge link.}
    \label{intamp}
\end{figure}

The longitudinal single spin asymmetry arises from the interference amplitudes between two gluon exchange diagram and the single photon exchange diagram know as the Primakoff process, as depicted in Fig.~\ref{intamp}.
It is easy to verify that the single spin asymmetry in exclusive $\pi^0$ production vanishes at the leading power. The  spin asymmetry has to be evaluated following the usual collinear expansion at the next to leading power.
Schematically, the spin dependent amplitude illustrated on the left side of the cut in Fig.~\ref{intamp} takes the form,
\begin{eqnarray} 
{\cal M}_L \propto \int d^2k_\perp \left \{ H(x,\xi, k_\perp)\Big|_{k_\perp=0}+\frac{\partial H(x,\xi, k_\perp)}{\partial k_\perp^\mu}\Big|_{k_\perp=0}k_\perp^\mu+... \right \} \, k_{\perp}\times  \Delta_\perp \lambda  \delta_{\lambda,\lambda'}    F_{1,4}^g 
\label{f14}
\end{eqnarray}
where  $ H(x,\xi, k_\perp)$ stands for the hard part. The leading power term drops out for the spin dependent contribution. 
$k_\perp$ is the relative transverse momentum of the  gluons entering the hard partonic part. Both $k_\perp$ and $\Delta_\perp$ will be set to zero in the hard part after performing the collinear expansion. Therefore, one may expect the vector product $\epsilon_\perp^{\gamma^*} \times k_\perp $ which respects the parity conservation shows up in the next to leading power hard part $\frac{\partial H}{ \partial k_\perp^\mu} k_\perp^\mu$. Carrying out $k_\perp$ integration in Eq.~\ref{f14}, the spin dependent amplitude  would be proportional to the vector structure,
 \begin{eqnarray} 
{\cal M}_L \propto \int d^2 k_\perp
 (\epsilon_\perp^{\gamma^*} \times k_\perp )(k_{\perp}\times  \Delta_\perp) \lambda  \delta_{\lambda,\lambda'}       F_{1,4}^g   \propto
 (\epsilon_\perp^{\gamma^*} \cdot   \Delta_\perp) \lambda  \delta_{\lambda,\lambda'}  \int d^2 k_\perp  \frac{1}{2}  k_\perp^2 F_{1,4}^g  
\end{eqnarray}
where the non-perturbative part is directly related to the gluon OAM~\cite{Lorce:2011kd,Hatta:2011ku,Lorce:2011ni,Hatta:2012cs}
in the forward limit $\Delta_\perp=0$.

On the other hand, $\pi^0$ can be exclusively produced in the Primakoff process~\cite{Primakoff:1951iae,Gasparian:2016oyl,Liping:2014wbp,Kaskulov:2011ab} as shown on the right side of the cut in Fig.~\ref{intamp}.  It has been recently discovered that in the small $x$ limit (or in the forward limit), the exchanged photon polarization is linearly polarized, with its polarization vector being parallel to its transverse momentum $\Delta_\perp^\nu$~\cite{Li:2019yzy,Li:2019sin,Xiao:2020ddm,Brandenburg:2021lnj,STAR:2019wlg,Zhou:2022twz}, in analogy to the QCD case~\cite{Metz:2011wb}. Apart from $\Delta_\perp^\nu$, another transverse vector  available in the hard part is virtual photon's transverse polarization vector. Therefore, to construct a scalar with these two transverse vectors,  the amplitude of the Primakoff process must be proportional to,
 \begin{eqnarray} 
{\cal M}_R^* \propto 
 \epsilon_\perp^{\gamma^*} \times  \Delta_\perp \, .    
\label{prim}
\end{eqnarray}
Combining the amplitudes ${\cal M}_R^*$ and ${\cal M}_L$, one gets,
\begin{eqnarray} 
{\cal M}_L {\cal M}_R^* \propto 
( \epsilon_\perp^{\gamma^*} \times  \Delta_\perp   )  (\epsilon_\perp^{\gamma^*} \cdot   \Delta_\perp) \lambda  \delta_{\lambda,\lambda'}  \int d^2 k_\perp  \frac{1}{2}  k_\perp^2 F_{1,4}^g \, .
\end{eqnarray}
The virtual photon's polarization is determined by the direction of the scattered electron's momentum $l'$.  The above angular correlation can be converted to  a $\sin 2\phi$ correlation at the full cross section level, where $\phi$ is the azimuthal angle between final state electron's transverse momentum $l_\perp'$ and $\Delta_\perp$. 
Two remarks now are in order.  The same $\sin 2\phi$ asymmetry could show up in the SIDIS process, that can give access to the specific TMDs~\cite{Li:2021mmi}. Second, it  often offers a novel way to study nucleon structure by making use of the Coulomb nuclear interference effects (see, for example~\cite{Hagiwara:2020juc}).

We now present the main analytical results in details. Let us start by reviewing the calculation of  the spin averaged cross section. The unpolarized amplitude involves the t-channel exchange of the spin dependent odderon, which is related to the dipole type gluon Sivers distribution. The unpolarized cross section can be expressed in terms of the Sivers function $f_{1T}^{\perp g}$~\cite{Boussarie:2019vmk},
\begin{eqnarray} 
\frac{d\sigma^{odderon}}{dt dQ^2 dx_B} \approx
\frac{\pi^5 \alpha_{em}^2 \alpha_s^2 f_\pi^2}{8x_B N_c^2M^2 Q^6}\left [1-y+\frac{y^2}{2}\right ]\! \left [ 
 \int_0^1 \! dz \frac{\phi_\pi(z)}{z(1-z)}\int \! d^2 k_\perp \frac{k_\perp^2xf_{1T}^{\perp g}(x,k_\perp^2) }{k_\perp^2+z(1-z)Q^2} \right ]^2  
\end{eqnarray}
where $\phi_\pi$ is the pion's leading twist distribution amplitude (DA) and $f_\pi=131$ MeV is the decay constant.  $z$ is the longitudinal momentum fraction of the virtual photon carried by quark. 

In the region where $\Delta_\perp$  is extremely small, exclusive $\pi^0$ production through the Primakoff process becomes important. The calculation of the Primakoff process was well formulated in terms of the pion distribution amplitude. The amplitude of the  Primakoff reads~\cite{Lepage:1980fj,Khodjamirian:1997tk}\footnote{Strictly speaking, in addition to the charge Form Factor (or the Dirac form factor, ${\cal{F}}$) the Pauli form factor $(F_2)$ contributes as well. Specifically, $F_2$ enters alongside ${\cal{F}}$ as $\, \sim {\cal{F}}-\tfrac{\xi^2}{1-\xi^2} F_2$. Since there is a relative suppression of $\xi^2$ between the two for the kinematics of our interest, we will refrain from explicitly writing this contribution from $F_2$. We are very grateful to Yoshitaka Hatta for bringing this point to our attention.},
\begin{eqnarray}
{\cal M}_{R} = \delta_{\lambda \lambda'} e^3  \sqrt{2}f_\pi \frac{\sqrt{1-\xi^2}}{1+\xi} \frac{ ( \epsilon_\perp^{\gamma^*} \times  \Delta_\perp   )  }{ x_B \Delta_\perp^2}  {\cal F}(t) \int_0^1 dz\frac{\phi_\pi(z)}{6z(1-z) }
\end{eqnarray}
where ${\cal F}(t)=1/\left (1+\frac{-t}{Q_0^2}\right )^2$ with $Q_0^2=0.71 \, {\textrm{GeV}}^2$, is the proton charge form factor. 
Correspondingly, the cross section of the Primakoff process is given by,
\begin{eqnarray}
\frac{d\sigma^{Pri}}{dt dQ^2 dx_B} \approx \frac{\alpha_{em}^4(2\pi) [1+(1-y)^2]f_\pi^2}{x_B Q^6  \Delta_\perp^2 } \frac{1-\xi}{1+\xi} {\cal F}^2(t)\left [ \int_0^1 \frac{dz}{6z(1-z)}\phi_\pi(z) \right ]^2 \, .
\end{eqnarray}
Due to its $1/\Delta_\perp^2$ behavior, the Primakoff process is the dominant  channel of the exclusive $\pi^0$ production  at low transverse momentum. Note that, as pointed in Ref.~\cite{Boussarie:2019vmk}, there is no interference between the spin dependent odderon and the Primakoff amplitude because the proton helicity flips with the odderon exchange while the proton helicity is conserved with one photon exchange. 

We now report the results of the amplitude that involves $F_{1,4}$. The calculation closely follows that outlined in Refs.~\cite{Ji:2016jgn,Bhattacharya:2022vvo}. There are a total of six diagrams contributing to the spin dependent amplitude, all of which vanish at the leading power. Two of them with the exchanged gluons attached to quark and anti-quark lines simultaneously survive at the next to leading power, whereas contributions from the other four diagrams drop out at the next to leading power after making $k_\perp$ expansion. We thus  explicitly verify that the single longitudinal spin asymmetry in exclusive $\pi^0$ production  is a twist-3 observable. The spin dependent amplitude is written as,
\begin{eqnarray}
{\cal M}_L= i\frac{g_s^2e f_\pi }{N_c\sqrt{2}} \lambda \delta_{\lambda \lambda'}  \frac{  \epsilon_\perp  \cdot  \Delta_\perp}{M^2Q^2} \frac{\xi^2(1+\xi)}{\sqrt{1-\xi^2}}\int_{-1}^1  dx \frac{F_{1,4}^{(1)} (x,\xi,\Delta_\perp) }{(x+\xi - i \epsilon)^2(x-\xi+i\epsilon)^2}   \int_0^1 dz \frac{\phi_\pi(z)}{z(1-z) } 
\end{eqnarray}
with $ F_{1,4}^{(1)} (x,\xi,\Delta_\perp)=\int d^2 k_\perp  k_\perp^2  F_{1,4} (x,\xi,\Delta_\perp,k_\perp)$. One notices that the third pole which could potentially break the factorization is absent in this amplitude, in contrast to the case of the diffractive di-jet production~\cite{Ji:2016jgn,Hatta:2016aoc,Bhattacharya:2022vvo}. Another possible source of twist-3 contribution is from $\Delta_\perp$ expansion. We have carried out an explicit calculation and found that the twist-3 contributions from  $\Delta_\perp$ expansion drop out for both unpolarized cross section and polarization dependent cross section.    Combining the amplitudes ${\cal M}_L$ and ${\cal M}_R^*$ with the leptonic tensor, we derive the full spin dependent cross section,
\begin{eqnarray}
\frac{d\Delta \sigma}{dt dQ^2 dx_B d\phi} &=& -\sin (2\phi)\frac{ \alpha_{em}^3 \alpha_s f_\pi^2 (1-y) \xi x_B {\cal F}(t)}{3Q^8   N_c} \nonumber \\
&& \hspace{0.5cm} \times \left[ \int_0^1 \! dz\frac{\phi_\pi(z)}{z(1-z)} \! \right]^2 {\rm Im}\! \left [  \int_{-1}^1  \! dx \frac{ F_{1,4}^{(1)} (x,\xi,\Delta_\perp)/M^2  }{(x+\xi - i \epsilon )^2(x-\xi+i\epsilon)^2}  \right ]   \label{cs}
\end{eqnarray}
where $\Delta \sigma=\sigma(\lambda=1)-\sigma(\lambda=-1)$.  The azimuthal angle $\phi$ is defined as the angle between the final state electron's transverse momentum and nucleon recoiled transvese momentum $\Delta_\perp$.  This is  the main result of our paper.

One notices that the real part of the gluon GTMD ${\rm Re}F_{1,4}$ is an odd function of $x$: ${\rm Re}F_{1,4}^{(1)} (x,\xi,\Delta_\perp)=-{\rm Re}F_{1,4}^{(1)} (-x,\xi,\Delta_\perp)$ simply because gluons are their own antiparticles. The integral involving the real part of $F_{1,4}^{(1)}$ thus vanishes,
\begin{eqnarray}
 {\rm Im} \int_{-1}^1  \! dx \frac{{\rm Re} F_{1,4}^{(1)} (x,\xi,\Delta_\perp)  }{(x+\xi-i\epsilon)^2(x-\xi+i\epsilon)^2} =0  \ .
\end{eqnarray}
This is also in accordance with the fact that exclusive $\pi^0$ production selects charge parity odd exchange as the C parity of virtual photon and $\pi^0$ are -1 and +1 respectively.  Two gluon exchange is ruled out by the C parity conservation. However, the imaginary part of $F_{1,4}$ is not necessarily an odd function of $x$, and thus could survive under $x$ integration.  One possible mechanism for generating an imaginary phase is through an additional gluon exchange which is encoded in the gauge link, in close analogy to the Sivers function case.  The $k_\perp$ moment of  the imaginary part of the gluon GTMD $F_{1,4}$ is thus related to a tri-gluon correlation function which represents a C-odd exchange.  However, the $\Delta_\perp$ expansion contribution is ruled out by   the C parity conservation as there is no such mechanism for generating the imaginary phase. 
Eq.~(\ref{cs}) clearly demonstrates that one can extract the imaginary part of  the characteristic azimuthal angular correlation $\sin 2\phi$ when the target is longitudinal polarized. 
Let us close this section with one final remark: the spin dependent cross section smoothly approaches a constant value when $\Delta_\perp \rightarrow 0$, whereas both the single and double spin asymmetries in  the diffractive di-jet production scale as $|\Delta_\perp|$ in the near forward limit.

We now numerically estimate the asymmetry with inputs from some toy model results. The azimuthal asymmetries, i.e., the average value of  $\sin (2\phi)$ that we are going to estimate numerically are defined as,
\begin{eqnarray}
\langle \sin(2\phi) \rangle &=&\frac{  \int \frac{d \Delta\sigma}{d {\cal P.S.}}\sin(2\phi) \ d {\cal P.S.} }
{\int \left [\frac{d \sigma^{Pri}}{d {\cal P.S.}}+\frac{d \sigma^{odderon}}{d {\cal P.S.}} \right ] d {\cal P.S.}} \, .
\end{eqnarray}
We start with calculating the unpolarized cross section. The only input we need to compute the cross section of the Primakoff process is the pion's distribution amplitude. As a first exploratory study, we neglect the scale dependence of the pion DA. For simplicity, we consider the asymptotic form for the pion's DA $\phi_\pi=6z(1-z)$. As for the dipole type gluon Sivers function i.e. the spin dependent odderon, there is currently no experimental constraint on it at all.  Several models for the gluon Sivers function are available in the  literature~\cite{Zhou:2013gsa,Szymanowski:2016mbq}. Here we simply parameterize the gluon Sivers function in terms of the unpolarized gluon distribution,
\begin{eqnarray}
f_{1T}^{\perp g}(x,k_\perp^2)={\cal C}\frac{x^{0.3}}{x_0}\frac{M}{|k_\perp|+\Lambda_{\rm QCD}}\frac{\delta_\perp^2}{2\pi}e^{-k_\perp^2/\delta_\perp^2}G(x)
\end{eqnarray}
with $x_0=0.01$. $G(x)$ the normal unpolarized gluon PDF. The width is  fixed to be  $\delta^2_\perp=0.53 \, \textrm{GeV}^2$.  The coefficient ${\cal C}=0.01$  is chosen 
 such that the QCD contribution to the cross section starts to dominate over the EM contribution when $\Delta_\perp > 150$ MeV.  Note that this parameterization does not satisfy the relation $\int d^2 k_\perp f_{1T}^{\perp g}(x,k_\perp^2)=0$ that holds at the tree level. As a matter of fact, such a relation quickly breaks down after performing the TMD evolution as shown in Ref.~\cite{Boer:2022njw}. 

To compute the spin dependent cross section, one has to model the Compton form factor that involves  $F_{1,4}^{(1)}$. Unfortunately, to the best of our knowledge, there is no model result available for the imaginary part of $F_{1,4}^{(1)}$. However, the imaginary part of $F_{1,4}^{(1)}$ is very unlikely larger than its real part. So we conjecture the following relation,
\begin{eqnarray}
{\rm Im}\!   \int_{0}^1  \! dx \frac{\frac{{\rm Im}F_{1,4}^{(1)}(x,\xi,\Delta_\perp)}{M^2}} {(x+\xi)^2(x-\xi+i\epsilon)^2} <
{\rm Im}\!   \int_0^1  \! dx \frac{\frac{{\rm Re} F_{1,4}^{(1)}(x,\xi,\Delta_\perp)}{M^2}} {(x+\xi)^2(x-\xi+i\epsilon)^2}  \approx \frac{\pi  }{2\xi}\frac{\partial}{\partial x} L(x,\xi)|_{x=\xi} \, .
\end{eqnarray}
In our numerical estimation, we replace the compton form factor with $\frac{\pi  }{2\xi}\frac{\partial}{\partial x} L(x,\xi)|_{x=\xi}$ to get a rough estimation of the upper limit of the asymmetry. Here we reconstruct the $\xi$-dependence for $xL_g(x,\xi)$ from it's PDF counterpart $xL_g(x)$ in accordance with the method of double distributions~\cite{Radyushkin:1998es,Radyushkin:2000uy}.   
To model the $x$-dependence for the OAM, we employ the Wandzura-Wilczek(WW) approximation~\cite{Hatta:2012cs},
\begin{eqnarray}
L_g(x)\approx x \int_x^1 \frac{dx'}{x'^2}\left [H_g(x')+E_g(x') \right ]-2x\int_x^1 \frac{dx'}{x'^2} \Delta G (x') + ...
\end{eqnarray}
In the forward limit, the gluon GPD $H_g$ is related to the normal gluon PDF $H_g(x)=x G (x)$. We use the JAM~\cite{Sato:2019yez,Ethier:2017zbq} gluon PDFs $xG(x)$ and $x\Delta G(x)$ as inputs.
The experimental studies of  the gluon GPD $E_g$ is rather sparse at the moment. One has to model it only with some theoretical guidance. A recent work has shown that $E_g$ grows very rapidly, and the ratio of $E_g(x)/H_g(x)$ approaches a constant in the limit $x\rightarrow 0$~\cite{Hatta:2022bxn}. In our numerical estimates, we simply assume $E_g(x)=cH_g(x)$ where the coefficient $c=-0.15$ is extracted from a  light-cone spectator model calculation~\cite{Tan:2023kbl}.

\begin{figure}
    \centering
\includegraphics[scale=0.6]{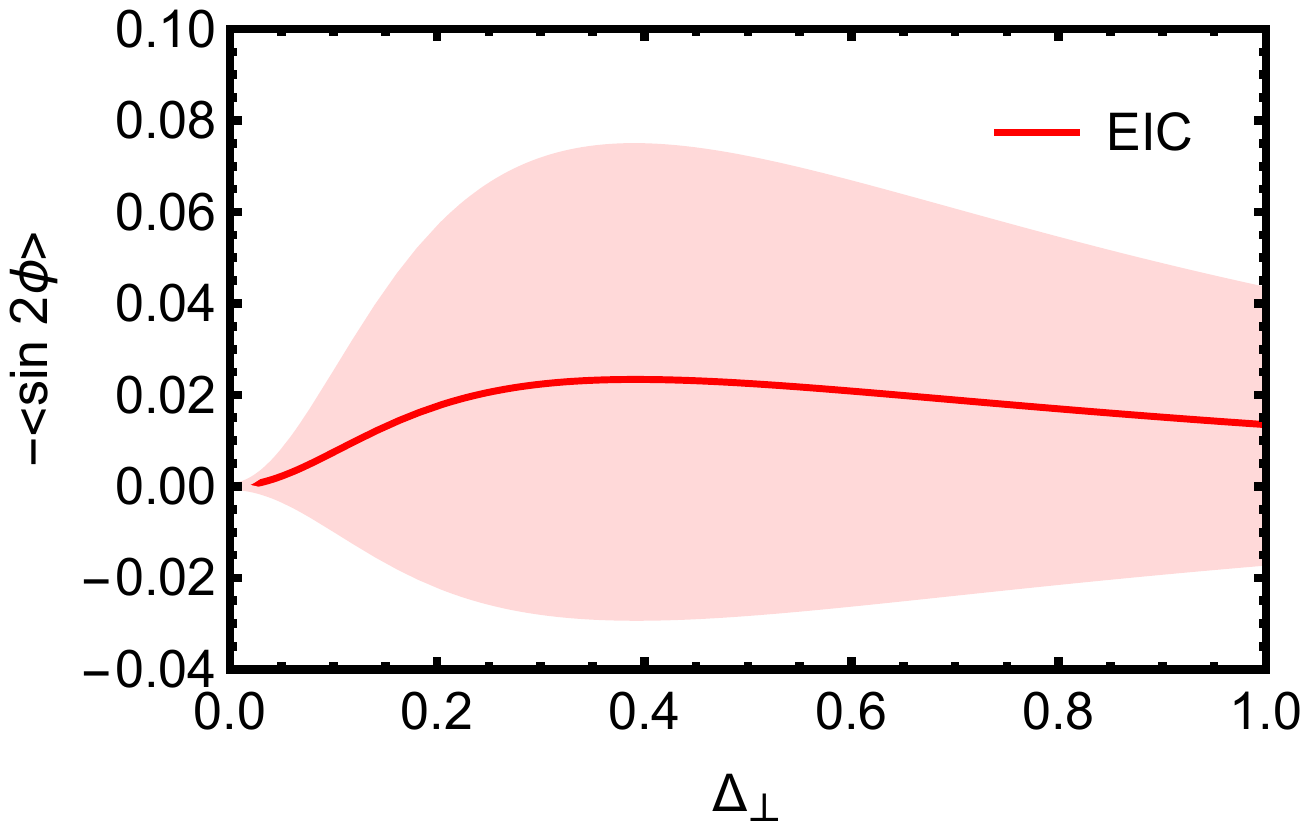}
\includegraphics[scale=0.6]{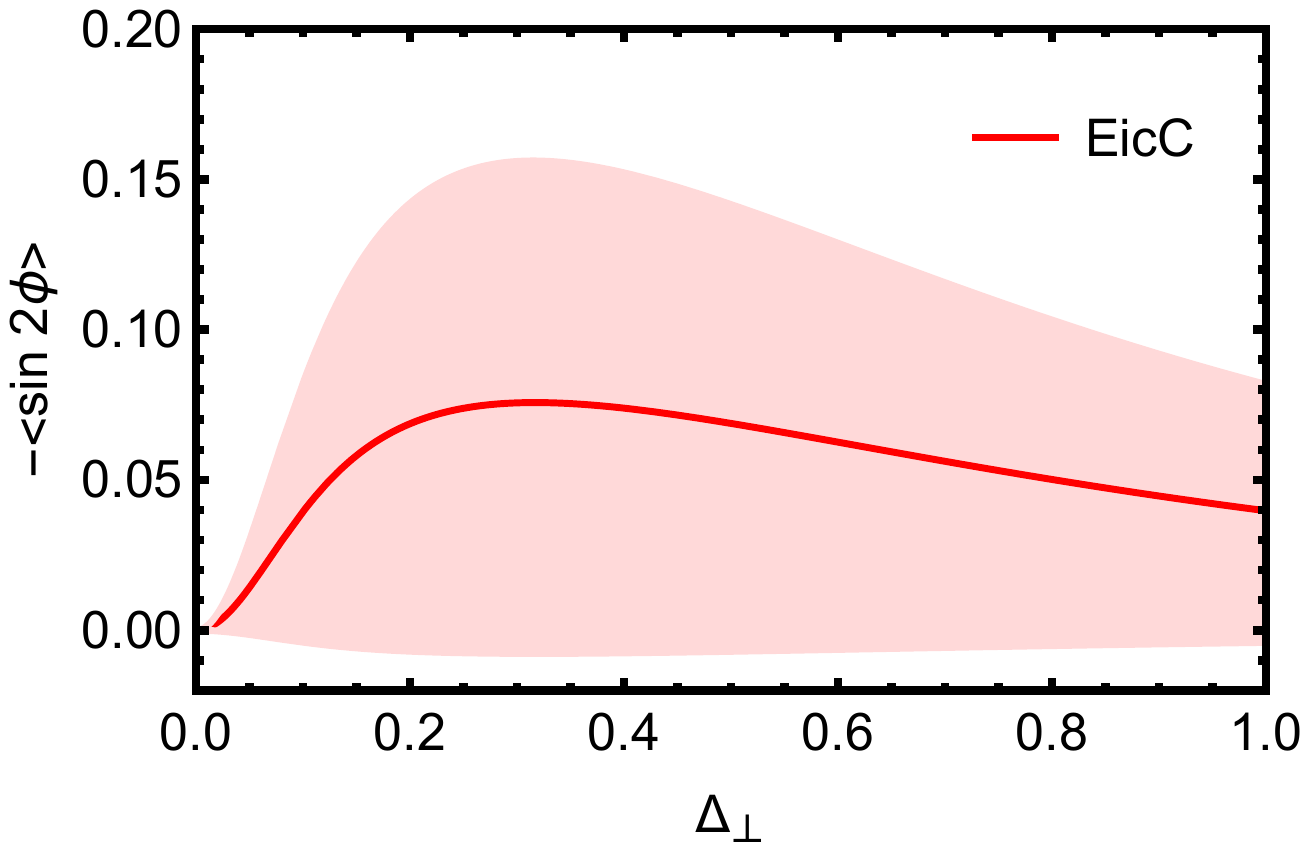}
    \caption{The averaged $\sin 2\phi$ azimuthal asymmetry computed at EIC and EicC energies is plotted as the function of $\Delta_\perp$.} 
    \label{0404 sin2phi eic}
\end{figure}

The  relevant kinematics are fixed to be $y=0.02$, $Q^2=10 \, {\textrm{GeV}}^2$, $\sqrt{S_{ep}}=$100 GeV for the EIC case and $y=0.5$, $Q^2=3 \, {\textrm{ GeV}}^2$, $\sqrt{S_{ep}}=$16 GeV
 for the EicC case. 
The asymmetry is displayed in Fig.~\ref{0404 sin2phi eic}  as the function of $\Delta_\perp$.  One can see that the asymmetries peak around $\Delta_\perp=300$ MeV. We also found that the asymmetry roughly scales as $\xi$. It is thus more promising to constrain the $F_{1,4}$  at relatively large $\xi$ region. Since there are huge uncertainties for the various gluon distributions involved in the calculation, our numerical estimates can only be considered as an exploratory study. 

\section{ Summary}
We propose extracting the imaginary part of the  gluon GTMD $F_{1,4}$ by measuring azimuthal angular correlation of $\sin 2\phi$ in exclusive $\pi^0$ production in  polarized $ep$ collisions.  Such a single target longitudinal spin asymmetry  arises from the interference effect between the QCD interaction and the Primakoff process. 
All the previous proposals of probing the gluon OAM from the single/double longitudinal spin dependent di-jet production rely on disentangling the gluon OAM distribution from the complicated compton form factor that may receive the contributions from both the imaginary part and  real part of $F_{1,4}$.   It is thus helpful for pining down the gluon OAM by first having an experimental constraint on the imaginary part of $F_{1,4}$ using the observable that we have studied. It would be interesting to carry out the measurement at the future EIC and EicC.

There are a number of directions along which the current work can be refined or extended. First, the single longitudinal spin asymmetry in exclusive $\pi^0$ may also receive the contributions from the different sources at the twist-3 level, for instance, the tri-gluon correlation functions which could play an important role.  Second, it would be interesting to explore the feasibility of extracting $F_{1,4}$ through the similar observable in the ultraperipheral polarized proton-heavy ion collisions. Moreover, the similar calculation can be performed for the quark channel that should be more relevant for the JLab experiment. Since quark distributions are in general neither even nor odd in $x$, the same $\sin 2\phi$ azimuthal asymmetry allows us to access the real and imaginary parts of $F_{1,4}$ simultaneously.   We  will address this issue  in a future publication~\cite{Bhattacharya:2023hbq}. 

\

{\it Acknowledgements: } We thank Yoshitaka Hatta, Andreas Metz, Zhun Lu, Xingbo Zhao and Siqi Xu for helpful discussions. This work has been supported by  the National Natural Science Foundation of China under Grant No.~12175118~(J.Z.), and the National Science Foundation under Contract No.~PHY-1516088. S.~B. has been supported by the U.S. Department of Energy under Contract No. DE-SC0012704, and also by  Laboratory Directed Research and Development (LDRD) funds from Brookhaven Science Associates. 

\

\bibliography{ref.bib}

\end{document}